\documentclass[aps, prb, reprint, superscriptaddress, floatfix]{revtex4-2}
\usepackage{graphicx}% Include figure files
\usepackage{dcolumn}% Align table columns on decimal point
\usepackage{bm}% bold math
\usepackage{physics}
\usepackage{amsfonts}
\usepackage{mathrsfs}
\usepackage{subfigure}
\usepackage[svgnames]{xcolor}
\usepackage[colorlinks=true,linkcolor=NavyBlue,anchorcolor=red,citecolor=NavyBlue, urlcolor=NavyBlue]{hyperref}
\usepackage{color}
\usepackage{xcolor}

\DeclareMathAlphabet{\pazocal}{OMS}{zplm}{m}{n}
\usepackage{amsmath}
\usepackage{amssymb}
\usepackage[T1]{fontenc}

\usepackage{color}
\usepackage{makecell}

\newcommand{\bs}[1]{\bm{#1}}
\newcommand{\mb}[1]{\bm{#1}}

\begin{document}

\title{Schwinger-Boson Mean-Field Study of the Anisotropic Kagome Antiferromagnet}

\author{Sankha Subhra Bakshi}
\affiliation{Department of Physics, University of Virginia, Charlottesville, Virginia, 22904, USA}

\author{Brandon B. Le}
\affiliation{Department of Physics, University of Virginia, Charlottesville, Virginia, 22904, USA}

\author{Seung-Hun Lee}
\affiliation{Department of Physics, University of Virginia, Charlottesville, Virginia, 22904, USA}

\author{Gia-Wei Chern}
\affiliation{Department of Physics, University of Virginia, Charlottesville, Virginia, 22904, USA}

\date{\today}

\begin{abstract}
We investigate the effect of spatial exchange anisotropy on the spin-$1/2$ kagome antiferromagnet using Schwinger-boson mean-field theory. The anisotropy is introduced by strengthening the Heisenberg exchange along one set of nearest-neighbor bonds relative to the other two, and is controlled by a parameter $\delta$ that measures the deviation from the isotropic limit. Incorporating the reduced lattice symmetry, we construct the corresponding projective-symmetry-group ans\"atze and focus on representative $0$- and $\pi$-flux states connected to the conventional $q=0$ and $\sqrt{3}\times\sqrt{3}$ kagome states. We find that anisotropy predominantly reconstructs the low-energy spinon sector, leading to a strong softening of the lowest spinon branch and a downward shift of the two-spinon continuum. At sufficiently large $\delta$, the spinon gap closes at ansatz-dependent values, signaling an instability toward spinon condensation and the onset of magnetic order. From the soft Bogoliubov eigenmodes, we reconstruct the associated incipient spin textures and show that the resulting magnetic orders are intrinsically anisotropic, with suppressed moments on strongly coupled bonds and enhanced moments on more weakly connected sites. These results provide a microscopic picture of how exchange anisotropy drives the transition from kagome spin-liquid states to magnetic order, and offer a framework for interpreting recent experiments on anisotropic kagome materials, particularly titanium-based spin-$1/2$ compounds.
\end{abstract}

\maketitle

\section{Introduction}

\label{sec:intro}

Quantum spin liquids (QSLs) are phases of frustrated magnets in which strong geometric frustration and quantum fluctuations suppress conventional symmetry-breaking order, leading instead to highly entangled states with fractionalized excitations and emergent gauge structures~\cite{Broholm2020,balents2010,savary2017,zhou2017}. The kagome lattice, composed of corner-sharing triangles, is a paradigmatic platform due to its large classical ground-state degeneracy~\cite{Chalker1992,Huse1992,Zhitomirsky2008,Chern2011} and strong quantum fluctuations~\cite{singh2007,yan2011}. Experimentally, this has been highlighted by kagome materials such as herbertsmithite ZnCu$_3$(OH)$_6$Cl$_2$~\cite{Mendels2010,Norman2016,Shores2005}, which shows no magnetic ordering down to temperatures far below the exchange scale and exhibits a broad continuum of spin excitations in neutron scattering~\cite{Helton2007,Han2012,Fu2015}. Related compounds, including volborthite, Zn-barlowite, and more recently Ti-based kagome magnets, further underscore the roles of anisotropy and disorder~\cite{Hiroi2001,Janson2016,Han2016,Feng2017,Fu2021,Thennakoon2025}. These systems establish kagome magnets as a central setting for QSL physics, while leaving open key questions about the nature and stability of their ground states.

From a theoretical perspective, the spin-$1/2$ Heisenberg antiferromagnet on the kagome lattice has been extensively investigated using a wide range of complementary analytical and numerical approaches, including parton constructions~\cite{sachdev1992,wang2006,ran2007,yan2011,messio2012}, variational Monte Carlo~\cite{Ferrari2023,kiese2023}, series expansions~\cite{singh2007}, exact diagonalization~\cite{Leung1993}, and density matrix renormalization group (DMRG)~\cite{yan2011,depenbrock2012,Iqbal2013,he2017,liao2017,mei2017,Zhu2019}. These studies have consistently pointed to a set of closely competing ground states, most prominently gapped $\mathbb{Z}_2$ spin liquids and gapless $U(1)$ Dirac spin liquids. While early large-scale DMRG calculations appeared to support a gapped ground state, subsequent work has highlighted the near-degeneracy of these phases and their pronounced sensitivity to microscopic details. Rather than selecting a single definitive ground state, the current understanding is better framed in terms of a manifold of proximate quantum phases residing within a narrow energy window~\cite{Hermele2008,Hu2015,changlani2018,Jiang2019}. This perspective naturally shifts the focus toward understanding how these competing spin-liquid states respond to perturbations, and in particular, how instabilities out of this manifold may give rise to distinct forms of magnetic order.

\begin{figure}[b!]
\includegraphics[width=0.99\columnwidth]{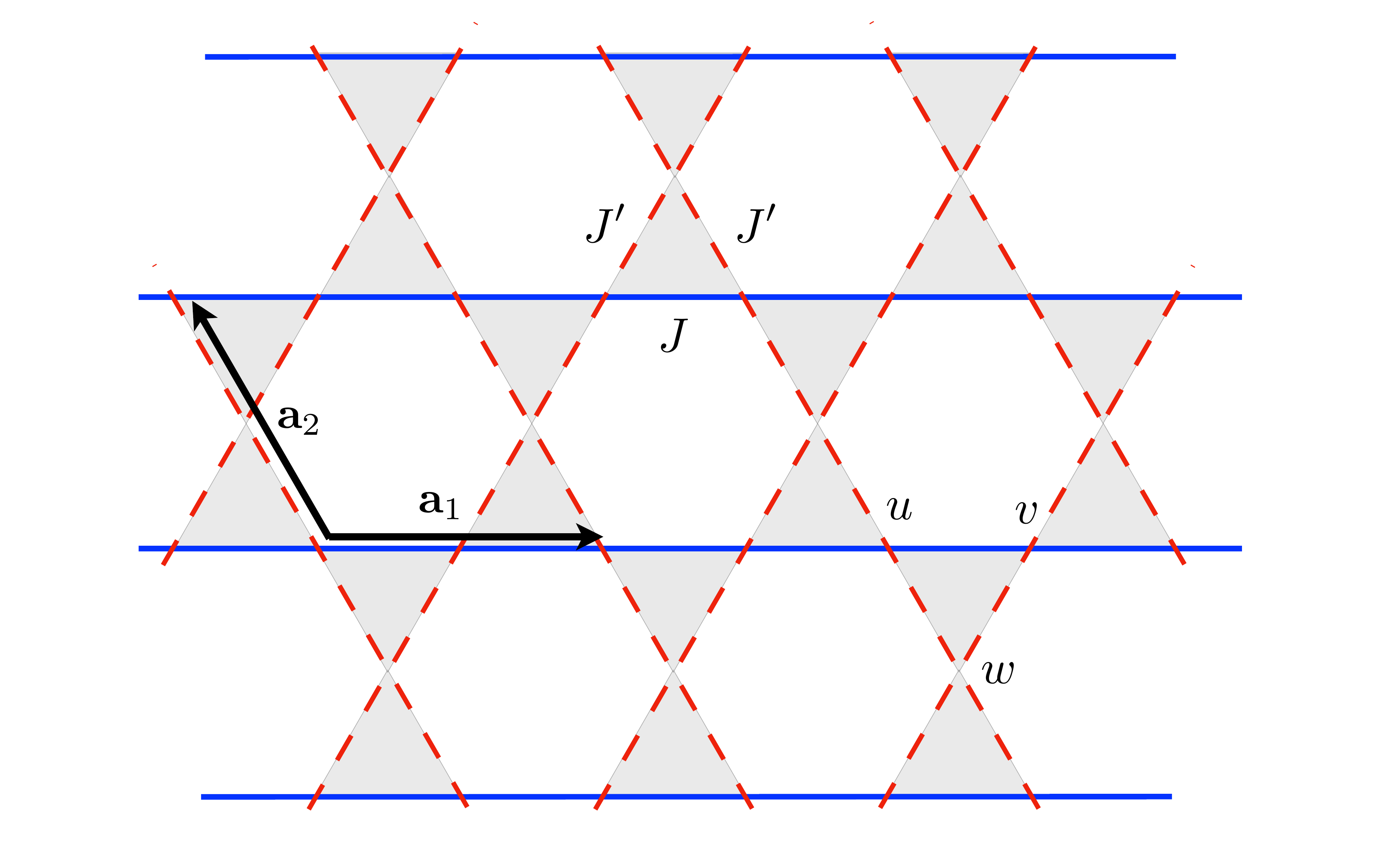}
\caption{
Schematic of the anisotropic kagome Heisenberg model with exchange couplings $J$ and $J'$. In the limit $J' \ll J$, the lattice can be viewed as weakly coupled spin chains connected by intermediate spins.}
\label{fig:kagome}
\end{figure}

A particularly instructive perturbation in this context is spatial anisotropy, introduced by allowing inequivalent nearest-neighbor exchanges $J$ and $J'$, as illustrated in Fig.~\ref{fig:kagome}. In the strongly anisotropic limit $J' \ll J$, the system can be viewed as a set of weakly coupled spin-$1/2$ chains connected through intermediate (“dangling”) spins. In this regime, controlled analytical approaches based on bosonization and renormalization-group methods have shown that the system develops magnetic order~\cite{stoudenmire2008,schnyder2008,zyuzin2012}. The interstitial spins order at a relatively high energy scale, forming a coplanar spiral with a small ordering wave vector, while the chain spins acquire a weaker, induced order through their coupling to these moments. Notably, this ordering pattern corresponds to an almost ferromagnetic alignment along the chains—i.e., a long-wavelength spiral—despite the underlying antiferromagnetic exchange $J$. This counterintuitive behavior reflects the effective interactions mediated by the interstitial spins and illustrates how anisotropy can selectively lift the degeneracy of the kagome spin-liquid manifold and stabilize qualitatively distinct magnetic states.

This leads to a noncoplanar and highly anisotropic magnetic structure, governed by a hierarchy of energy scales associated with interstitial and chain degrees of freedom~\cite{stoudenmire2008,schnyder2008,zyuzin2012}. While this analysis provides valuable insight into the quasi-one-dimensional limit, it leaves open how this ordered state evolves as the anisotropy is reduced and the system approaches the isotropic kagome point, where spin liquid behavior is widely expected.
Motivated by this, it is natural to seek a systematic description of the anisotropic kagome model across a broad range of exchange anisotropy, with the goal of understanding the stability of candidate spin liquid states. Numerical approaches such as DMRG, while powerful, become increasingly demanding when exploring extended phase diagrams and closely competing states. This motivates the use of complementary approaches that can efficiently access both disordered and ordered regimes.

In this work, we study the anisotropic kagome Heisenberg antiferromagnet using the Schwinger boson formalism, with an emphasis on the evolution of the ground state as a function of $J'/J$. We first carry out a projective symmetry group (PSG)~\cite{Wen2002A,Wen2002B} analysis to classify the symmetric spin liquid Ans\"atze allowed by the reduced lattice symmetry. Focusing on representative zero- and $\pi$-flux states, we then perform a systematic Schwinger boson mean-field analysis across the anisotropy range. We find that the spinon gap decreases with increasing anisotropy and closes at a finite $J'/J$, signaling an instability of the spin liquid toward magnetic order. This provides a coherent picture connecting the spin liquid regime near the isotropic limit to the magnetically ordered phase in the strongly anisotropic regime, and complements previous studies that primarily focused on limiting cases.

The remainder of this paper is organized as follows. In Sec.~II, we introduce the Schwinger-boson mean-field formalism and detail its implementation for the anisotropic kagome lattice, including the projective symmetry group (PSG) analysis, the construction of self-consistent saddle points, and the computation of the dynamical structure factor. In Sec.~III, we present our main results, focusing on the evolution of the candidate spin-liquid states under exchange anisotropy and the corresponding changes in their excitation spectra. In Sec.~IV, we analyze the magnetic orders that emerge in proximity to these spin-liquid phases, obtained from the condensation of the soft spinon modes, and characterize their ordering wave vectors and spin structures. Finally, Sec.~V summarizes our conclusions and discusses possible future directions.

\section{Schwinger boson mean-field theory}

We consider the nearest-neighbor antiferromagnetic Heisenberg model on the kagome lattice,
\begin{equation}
    \hat{\mathcal{H}} = \sum_{\langle ij\rangle} J_{ij}\, \hat{\bf S}_i\cdot \hat{\bf S}_j ,
    \label{eq:Hspin}
\end{equation}
where spatial anisotropy is introduced by taking $J_{ij}=J$ on bonds parallel to the primitive vector ${\bf a}_1$ and $J_{ij}=J'$ on the two symmetry-related directions ${\bf a}_2$ and ${\bf a}_1-{\bf a}_2$; see Fig.~\ref{fig:kagome}. The isotropic kagome limit is recovered at $J'=J$, and we define $\delta=1-J'/J$ as a convenient measure of anisotropy.

To investigate the resulting quantum phases and their instabilities, we employ the Schwinger-boson mean-field framework. This approach offers a unified description of both quantum disordered and magnetically ordered states. In particular, gapped $\mathbb{Z}_2$ spin liquids arise naturally at the mean-field level, while magnetic order emerges through the condensation of bosonic spinons, providing a direct route to tracking instabilities of the spin-liquid manifold. This capability is especially advantageous in the present setting, where exchange anisotropy is expected to destabilize spin-liquid phases and drive the system toward magnetic order. By contrast, fermionic parton constructions are more naturally suited to describing gapless $U(1)$ spin liquids. Here, we instead focus on the stability of gapped states and their evolution toward ordered phases, for which the Schwinger-boson formulation is particularly well suited.

This choice is further motivated by the broad success of the Schwinger-boson framework in describing kagome antiferromagnets under a variety of perturbations~\cite{messio2010,huh2010,messio2013,Dodds2013,Ern2017,rossi2023}. In such settings, it has proven to be a versatile platform for exploring the interplay between competing quantum disordered and ordered phases. For example, Dzyaloshinskii–Moriya interactions are known to substantially reshape the phase diagram and dynamical response within this approach. Likewise, further-neighbor couplings and more intricate exchange patterns---relevant to materials such as volborthite---have been shown to stabilize a variety of competing states within Schwinger-boson mean-field theory~\cite{Ern2017}. Beyond static properties, extensions that incorporate gauge fluctuations and topological excitations can yield a more realistic description of experimental observables. In particular, the inclusion of vison excitations in $\mathbb{Z}_2$ spin liquids has been demonstrated to qualitatively reproduce key features of the measured dynamic structure factor in herbertsmithite~\cite{Punk2014}, underscoring the ability of suitably extended Schwinger-boson theories to bridge microscopic models and experimental signatures.

Within this formalism, the spin operators are represented as
\begin{equation}
    \hat{\bf S}_i = \frac12 \hat b_{i,\alpha}^{\dagger} {\boldsymbol\sigma}_{\alpha\beta} \hat b_{i,\beta},
    \qquad \hat n_i=\hat b^\dagger_{i,\alpha}\hat b_{i,\alpha}=2S ,
    \label{eq:schwinger}
\end{equation}
with the local boson-number constraint enforced on average via Lagrange multipliers. Introducing the singlet pairing and hopping operators
\begin{eqnarray}
	\label{eq:AB_def}
	\hat A_{ij}=\frac{1}{2}\epsilon_{\alpha\beta}\hat b_{i,\alpha}\hat b_{j,\beta}, \qquad 
	\hat B_{ij}=\frac{1}{2}\hat b_{i,\alpha}^{\dagger}\hat b_{j,\alpha},
\end{eqnarray}
the exchange interaction can be written as
\begin{equation}
    \hat{\bf S}_i\cdot\hat{\bf S}_j = :\hat B_{ij}^{\dagger}\hat B_{ij}: - \hat A_{ij}^{\dagger}\hat A_{ij}.
\end{equation}
The corresponding mean-field amplitudes are $\mathcal A_{ij}=\langle\hat A_{ij}\rangle$ and $\mathcal B_{ij}=\langle\hat B_{ij}\rangle$, which respectively encode singlet pairing correlations and bosonic hopping processes.

Retaining both channels, the quadratic mean-field Hamiltonian reads
\begin{eqnarray}
    \hat{\mathcal{H}}_{\rm MF} &=& \sum_{\langle ij\rangle} J_{ij} \Big[ \mathcal B_{ij}\hat B_{ij}^{\dagger} +\mathcal B_{ij}^{*}\hat B_{ij}
        -\mathcal A_{ij}\hat A_{ij}^{\dagger} -\mathcal A_{ij}^{*}\hat A_{ij} \Big] \nonumber\\
    &+&  \sum_{\langle ij\rangle}J_{ij} \left( |\mathcal A_{ij}|^2-|\mathcal B_{ij}|^2 \right) 
    + \sum_i\lambda_i(\hat n_i-2S).
    \label{eq:HMF_real}
\end{eqnarray}
Physically, $\mathcal A_{ij}$ favors singlet formation and quantum disorder, while $\mathcal B_{ij}$ controls spinon hopping and dispersion.

The mean-field parameters $\{\mathcal A_{ij},\mathcal B_{ij},\lambda_i\}$ are determined self-consistently by minimizing the ground-state energy (or free energy at finite temperature) subject to the boson-number constraint. In practice, one diagonalizes $\hat{\mathcal H}_{\rm MF}$, evaluates the expectation values $\langle \hat A_{ij}\rangle$, $\langle \hat B_{ij}\rangle$, and $\langle \hat n_i\rangle$, and iterates until convergence.

\subsection{Projective symmetry group analysis}
\label{sec:psg}

We now outline the classification of symmetric Schwinger-boson mean-field ans\"atze within the projective symmetry group (PSG) framework~\cite{wang2006}; further details of the derivation are provided in Appendix~\ref{app:psg}. The presence of spatial anisotropy lowers the point-group symmetry relative to the isotropic kagome lattice, thereby modifying the PSG constraints and enlarging the set of symmetry-allowed ans\"atze.

We describe the kagome lattice as a triangular Bravais lattice with primitive vectors $\mathbf{a}_1,\mathbf{a}_2$ and a three-site basis $p=u,v,w$. A site is labeled by $(r_1,r_2)_p$. The symmetry group of the anisotropic model is generated by translations $T_1,T_2$, a reflection $\sigma$, and a $\pi$-rotation $R_\pi$, satisfying
\begin{eqnarray*}
	& & T_1T_2 = T_2T_1, \\
	& & \sigma^2 = 1, \quad R_\pi^2 = 1, \quad \sigma R_\pi = R_\pi \sigma, \\
	& & \sigma T_1 \sigma^{-1} = T_2, \quad \sigma T_2 \sigma^{-1} = T_1, \\
	& & R_\pi T_i R_\pi^{-1} = T_i^{-1}.
\end{eqnarray*}

In the Schwinger-boson representation, symmetry operations act projectively:
\begin{equation}
	b_{i\alpha} \rightarrow e^{i\phi_X(i)} b_{X(i)\alpha},
\end{equation}
with a site-dependent gauge phase $\phi_X(i)$. For $\mathbb{Z}_2$ spin liquids, PSG relations are defined up to a sign.

After gauge fixing, the algebraic PSG solutions can be parameterized as
\begin{align}
	&  \phi_{T_1,p}(r_1,r_2) = 0, \qquad \phi_{T_2,p}(r_1,r_2) = p_1 \pi r_1, \nonumber \\
	&  \phi_{\sigma,p}(r_1,r_2) = \frac{p_2\pi}{2} + p_1 \pi r_1 r_2, \nonumber \\
	&  \phi_{R_\pi,u}(r_1,r_2) = \frac{(p_3+p_4)\pi}{2} + p_4 \pi (r_1+r_2), \\
	&  \phi_{R_\pi,v}(r_1,r_2) = \frac{(p_3+p_4)\pi}{2} + p_1\pi + p_4\pi r_1 + (p_4+p_1)\pi r_2, \nonumber \\
	&  \phi_{R_\pi,w}(r_1,r_2) = \frac{(p_3+p_1)\pi}{2} + p_4\pi r_1 + (p_4+p_1)\pi r_2, \nonumber
\end{align}
with $p_1,p_2,p_3,p_4 \in \{0,1\}$, giving $16$ algebraic PSGs.

\begin{table}[b]
\caption{Classification of nearest-neighbor Schwinger-boson ans\"atze
on the anisotropic kagome lattice. The PSG is labeled by
$(p_1,p_3,p_4)$ with $p_2=1$.}
\label{tab:psg_ansatz}
\begin{ruledtabular}
\begin{tabular}{cccccc}
$p_1$ & $p_3$ & $p_4$ & flux & $e^{i \phi_{uv}}$ & $e^{i \phi_{uw}} = e^{i \phi_{vw}}$ \\
\hline
0 & 0 & 0 & $0$   & $+$ & $+$ \\
0 & 0 & 1 & $0$   & $+$ & $-$ \\
0 & 1 & 0 & $0$   & $-$ & $+$ \\
0 & 1 & 1 & $0$   & $-$ & $-$ \\
1 & 0 & 0 & $\pi$ & $+$ & $+$ \\
1 & 0 & 1 & $\pi$ & $+$ & $-$ \\
1 & 1 & 0 & $\pi$ & $-$ & $+$ \\
1 & 1 & 1 & $\pi$ & $-$ & $-$ \\
\end{tabular}
\end{ruledtabular}
\end{table}

Imposing compatibility with the nearest-neighbor pairing ansatz further constrains these solutions. For example, considering the $u$--$v$ bond within a unit cell and its transformation under reflection $\sigma$, one finds
\begin{equation}
	\mathcal{A}_{uv} = - e^{i (\phi_{\sigma,u}+\phi_{\sigma,v}) } \mathcal{A}_{uv},
\end{equation}
which requires $e^{i[\phi_{\sigma,u}(0,0)+\phi_{\sigma,v}(0,0)]}=-1$. This fixes $p_2=1$, while no further constraints arise. Therefore the allowed PSGs are labeled by
\begin{equation}
	p_2=1, \qquad p_1,p_3,p_4 \in \{0,1\},
\end{equation}
yielding $8$ distinct nearest-neighbor ans\"atze.
The parameter $p_1$ determines the gauge flux per unit cell, distinguishing $0$-flux and $\pi$-flux sectors. The parameters $p_3$ and $p_4$ encode the relative gauge structure under lattice rotations and distinguish different symmetry implementations within each flux sector.

The gauge flux through a closed loop is defined as the phase accumulated by the product of pairing amplitudes along the loop. For an elementary triangle $(i,j,k)$,
\begin{equation}
    \Phi_{\triangle} = \arg \left( \mathcal A_{ij}\mathcal A_{jk}\mathcal A_{ki} \right),
\end{equation}
which is gauge invariant modulo $2\pi$. Similarly, one may define the flux through larger loops such as hexagons. In the nearest-neighbor ansatz considered here, the flux sector (0 or $\pi$) can be unambiguously characterized by these elementary loops.

To construct explicit ans\"atze, we parametrize the pairing amplitudes on symmetry-inequivalent bonds as
\begin{eqnarray}
	& \mathcal{A}_{uv} = \mathcal{A}_1 e^{i \phi_{uv} }, \qquad & \mathcal{A}_{uw} = \mathcal{A}_{vw} = \mathcal{A}_2 e^{i \phi_{uw}}, \\
	& \mathcal{B}_{uv} = \mathcal{B}_1, \qquad & \mathcal{B}_{uw} = \mathcal{B}_{vw} = \mathcal{B}_2.
\end{eqnarray}

The eight resulting ans\"atze are summarized in Table~\ref{tab:psg_ansatz}. It is useful to relate these states to the well-known PSG classification of the isotropic kagome lattice. In that case, higher lattice symmetry constrains the ansatz more strongly, leading to four distinct nearest-neighbor PSG solutions (two in the $0$-flux sector and two in the $\pi$-flux sector). 

In the present anisotropic model, the reduced symmetry splits these into eight states. More precisely, each isotropic PSG state corresponds to two anisotropic PSGs distinguished by $(p_3,p_4)$, reflecting the fact that certain symmetry operations that were previously equivalent are no longer related. Thus, the anisotropic PSG classification can be viewed as a symmetry-resolved refinement of the isotropic one. Each PSG class corresponds to a distinct $\mathbb{Z}_2$ spin-liquid ansatz characterized by its gauge flux and bond-dependent sign structure. Energetic considerations are required to determine which of these candidate states is realized for given couplings $(J,J')$.

\subsection{Mean-field solutions}

We focus here on the simplest $0$-flux and $\pi$-flux solutions listed in the PSG classification. These two classes correspond, respectively, to the $q=0$ and $\sqrt{3}\times\sqrt{3}$ patterns familiar from the isotropic kagome problem, but in the present anisotropic setting they acquire additional structure due to the inequivalence between $J$ and $J'$ bonds. In particular, the reduced lattice symmetry allows for two independent on-site constraints, leading to distinct Lagrange multipliers on the inequivalent sublattices: $\lambda_u=\lambda_v\equiv\lambda_1$ and $\lambda_w\equiv\lambda_2$. Physically, this reflects the fact that the boson density (and hence the local constraint enforcing $\langle \hat n_i\rangle=2S$) may differ between the $w$ sublattice and the $u,v$ sublattices at the mean-field level.

To diagonalize the quadratic Hamiltonian, we first perform a Fourier transform of the boson operators, $\hat b_{m,s,\alpha} = N^{-1/2}\sum_{\bf k} \hat b_{{\bf k},s,\alpha} e^{i{\bf k}\cdot({\bf R}_m+{\bf r}_s)}$, where $s=u,v,w$ labels the three sublattices in the unit cell. We further introduce the shorthand $k_a={\bf k}\cdot{\bf e}_a$ associated with the bond directions. In momentum space, the Hamiltonian naturally takes a Bogoliubov–de Gennes form due to the presence of pairing amplitudes $\mathcal A_{ij}$, which mix creation and annihilation operators. It is therefore convenient to introduce the Nambu spinor
\begin{equation}
    \hat\Psi_{\bf k} = (
    \hat b_{{\bf k},u,\uparrow},
    \hat b_{{\bf k},v,\uparrow},
    \hat b_{{\bf k},w,\uparrow},
    \hat b^\dagger_{-{\bf k},u,\downarrow},
    \hat b^\dagger_{-{\bf k},v,\downarrow},
    \hat b^\dagger_{-{\bf k},w,\downarrow}
    )^T ,
\end{equation}
in which particle and hole sectors are treated on equal footing. In this basis, the mean-field Hamiltonian becomes
\begin{equation}
    \hat{\mathcal{H}}_{\rm MF} = \sum_{\bf k} \hat\Psi_{\bf k}^{\dagger} D_{\bf k} \hat\Psi_{\bf k} +E_c ,
    \label{eq:HMF_k}
\end{equation}
where $E_c$ collects all contributions arising from the mean-field decoupling,
\begin{align}
    E_c &=     2NJ(\mathcal A_1^2-\mathcal B_1^2) +4NJ'(\mathcal A_2^2-\mathcal B_2^2)     \nonumber\\
    &\quad -N(2\lambda_1+\lambda_2)(1+2S).
    \label{eq:Ec}
\end{align}
The first line represents the energy cost (or gain) associated with bond pairing ($\mathcal A$) and hopping ($\mathcal B$) amplitudes on $J$ and $J'$ links, while the second line enforces the boson number constraint on average.

The Bogoliubov matrix $D_{\bf k}$ encodes the full quadratic structure:
\begin{equation}
    D_{\bf k} = \begin{pmatrix}
        R_{\bf k} & e^{i\phi/2}P_{\bf k}\\
        e^{-i\phi/2}P_{\bf k}^{T} & R_{\bf k}
    \end{pmatrix} + \Lambda ,
    \label{eq:Dk}
\end{equation}
where $\Lambda= \operatorname{diag}
    (\lambda_1,\lambda_1,\lambda_2,
     \lambda_1,\lambda_1,\lambda_2)$ shifts the onsite energies and enforces the constraint. The matrix structure clearly separates particle-number-conserving processes ($R_{\bf k}$) from pairing processes ($P_{\bf k}$). Explicitly,
\begin{align}
	R_{\bf k} &= \begin{pmatrix} 0 & J\mathcal B_1\cos k_1 & J'\mathcal B_2\cos k_3 \\
	J\mathcal B_1\cos k_1 & 0 & J'\mathcal B_2\cos k_2 \\
	J'\mathcal B_2\cos k_3 & J'\mathcal B_2\cos k_2 & 0 
	\end{pmatrix},
	\label{eq:Rk_matrix}
\end{align}
describes spinon hopping between sublattices. The cosine structure originates from summing over bonds related by inversion, and the anisotropy between $J$ and $J'$ is directly reflected in the different amplitudes and momentum dependencies. Similarly,
\begin{widetext}
\begin{align}
	P_{\bf k} &= \begin{pmatrix}
	0 & -J\mathcal A_1\cos\!\left(k_1-\frac{\phi}{2}\right) & J'\mathcal A_2\cos\!\left(k_3+\frac{\phi}{2}\right) \\
	J\mathcal A_1\cos\!\left(k_1+\frac{\phi}{2}\right) & 0 & -J'\mathcal A_2\cos\!\left(k_2-\frac{\phi}{2}\right) \\
	-J'\mathcal A_2\cos\!\left(k_3-\frac{\phi}{2}\right) & J'\mathcal A_2\cos\!\left(k_2+\frac{\phi}{2}\right) & 0
	\end{pmatrix},
	\label{eq:Pk_matrix}
\end{align}
\end{widetext}
encodes singlet pairing of bosons. The phase $\phi$ captures the gauge-invariant flux threading elementary loops, distinguishing the $0$- and $\pi$-flux ans\"atze. The momentum shifts $\pm\phi/2$ reflect how the pairing amplitude transforms under lattice translations in the presence of flux.

Since Eq.~\eqref{eq:HMF_k} describes a bosonic quadratic Hamiltonian, it must be diagonalized by a \emph{paraunitary} Bogoliubov transformation $\hat\Psi_{\bf k}=M_{\bf k}\hat\Gamma_{\bf k}$ satisfying $M_{\bf k}^{\dagger}\tau_3M_{\bf k}=\tau_3$, where $\tau_3=\operatorname{diag}(1,1,1,-1,-1,-1)$. This condition ensures preservation of bosonic commutation relations. The excitation spectrum is obtained from the generalized eigenvalue problem
\begin{equation}
    \tau_3D_{\bf k}{\bf v}_{n{\bf k}} =  \omega_n({\bf k}){\bf v}_{n{\bf k}},     \qquad
    {\bf v}_{n{\bf k}}^{\dagger}\tau_3{\bf v}_{n{\bf k}}>0 ,
    \label{eq:bosonic_eig}
\end{equation}
whose positive-norm solutions yield the physical spinon bands $\omega_n({\bf k})$. A physically stable mean-field solution requires all $\omega_n({\bf k})$ to be real and non-negative; otherwise the saddle point is unstable.

For later use, we decompose the Bogoliubov matrix as
\begin{equation}
    M_{\bf k} =
    \begin{pmatrix}
        U_{\bf k} & X_{\bf k}\\
        V_{\bf k} & Y_{\bf k}
    \end{pmatrix},
    \label{eq:M_blocks}
\end{equation}
where each block is a $3\times3$ matrix in sublattice and band space. These matrices encode the coherence factors of the Bogoliubov quasiparticles and will directly enter physical observables.

The six mean-field variational parameters $\bm x = \{\mathcal A_1,\mathcal A_2,\mathcal B_1,\mathcal B_2,\lambda_1,\lambda_2\}$ are determined by minimizing the ground-state energy, $\partial E_{\rm MF}/\partial x_a=0$. Equivalently, one enforces the self-consistency conditions $\mathcal A_{ij}=\langle\hat A_{ij}\rangle$, $\mathcal B_{ij}=\langle\hat B_{ij}\rangle$, and $\langle\hat n_i\rangle=2S$. At zero temperature, these expectation values are computed using the Bogoliubov vacuum defined by Eq.~\eqref{eq:bosonic_eig}. In practice, this leads to a set of coupled nonlinear equations that must be solved iteratively, with the spinon spectrum feeding back into the bond amplitudes until convergence is reached.

\subsection{Dynamical Structure Factor}

The dynamical structure factor, directly measurable in inelastic neutron scattering, is defined as
\begin{equation}
    \mathcal{S}({\bf k},\omega) = \frac{1}{3N} \sum_{i,j} e^{-i{\bf k}\cdot({\bf r}_i-{\bf r}_j)} 
    \int_{-\infty}^{\infty}dt\, e^{i\omega t} \langle \hat{\bf S}_i(t)\cdot\hat{\bf S}_j(0) \rangle .
    \label{eq:DSF_def}
\end{equation}
Within the Schwinger boson framework, the spin operator is bilinear in bosons, implying that $\mathcal S({\bf k},\omega)$ probes two-spinon excitations. At zero temperature, the Bogoliubov vacuum contains no pre-existing spinons, and therefore the leading contribution arises from creating a pair of spinons with total momentum ${\bf k}$.

As a result, the structure factor takes the form
\begin{align}
    \mathcal{S}({\bf k},\omega) &= \frac{1}{12N} \sum_{\bf q} \sum_{r,s,m,n} \mathcal D^{r,s,m,n}_{{\bf k},{\bf q}}     \nonumber\\
    &\quad\times \delta[ \omega-\omega_m(-{\bf q})  -\omega_n({\bf k}+{\bf q})     ] ,
    \label{eq:DSF_T0}
\end{align}
which describes a continuum of excitations corresponding to all possible ways of partitioning the total momentum between two spinons. The delta function enforces energy conservation, and the resulting spectrum is typically broad, reflecting the fractionalized nature of the excitations.

The intensity is modulated by the coherence factor~\cite{halimeh2019}
\begin{equation}
    \mathcal D^{r,s,m,n}_{{\bf k},{\bf q}} = \mathcal M^{r,m,n}_{{\bf k},{\bf q}} \left( \mathcal M^{s,m,n}_{{\bf k},{\bf q}} \right)^* ,
    \label{eq:D_coherence}
\end{equation}
where
\begin{equation}
    \mathcal M^{r,m,n}_{{\bf k},{\bf q}} = V_{rm}(-{\bf q}) Y^*_{rn}({\bf k}+{\bf q}) -  X^*_{rm}(-{\bf q})  U_{rn}({\bf k}+{\bf q}) .
    \label{eq:M_coherence}
\end{equation}
These factors encode the internal structure of the spinon wavefunctions and determine how strongly different two-spinon states couple to the physical spin operator. In particular, they contain interference effects between particle-like and hole-like components of the Bogoliubov quasiparticles, and are therefore highly sensitive to the underlying mean-field ansatz (e.g., $0$ vs.\ $\pi$ flux). 

Consequently, even when the spinon dispersion $\omega_n({\bf k})$ is similar between different ans\"atze, the dynamical structure factor can display qualitatively distinct features, making it a powerful diagnostic for distinguishing competing quantum spin liquid states.

\section{Symmetric spin liquid phases}

In the following we present the zero-temperature Schwinger-boson mean-field results for the anisotropic kagome antiferromagnet. Throughout this work we fix $S=0.2$. Within SBMFT, this choice stabilizes a gapped spin-liquid saddle point for the isotropic kagome model and avoids the premature onset of magnetic order that can occur at larger values of $S$. Starting from the isotropic limit, we vary spatial anisotropy through $\delta = 1 - J'/J$ and systematically track the evolution of the self-consistent mean-field parameters, the spinon excitation gap, the spinon dispersions, and the zero-temperature dynamical structure factor.
%===========================================
\begin{figure*}[t!]
\includegraphics[width=8.8cm,height=8cm]{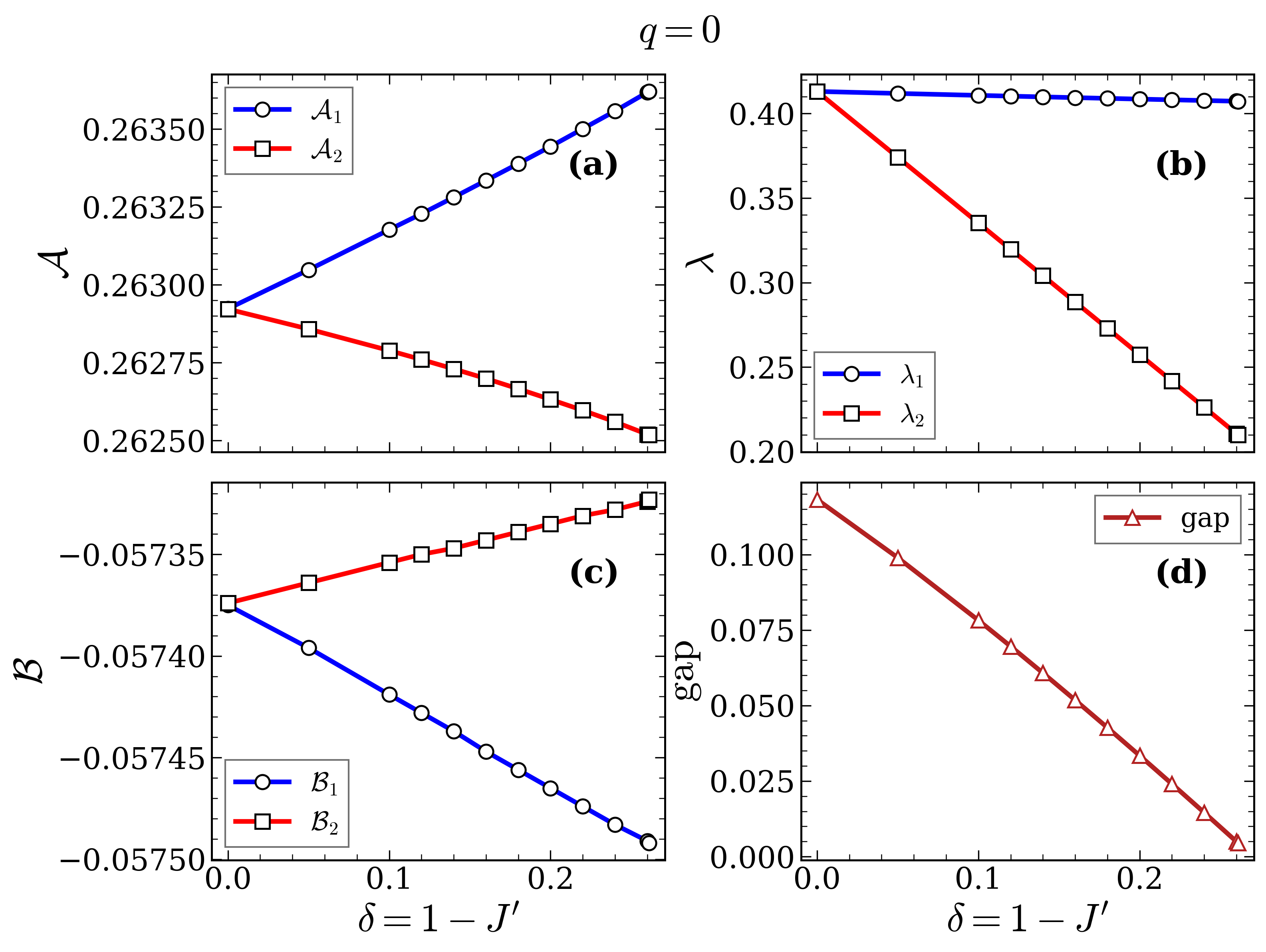}
\includegraphics[width=8.8cm,height=8cm]{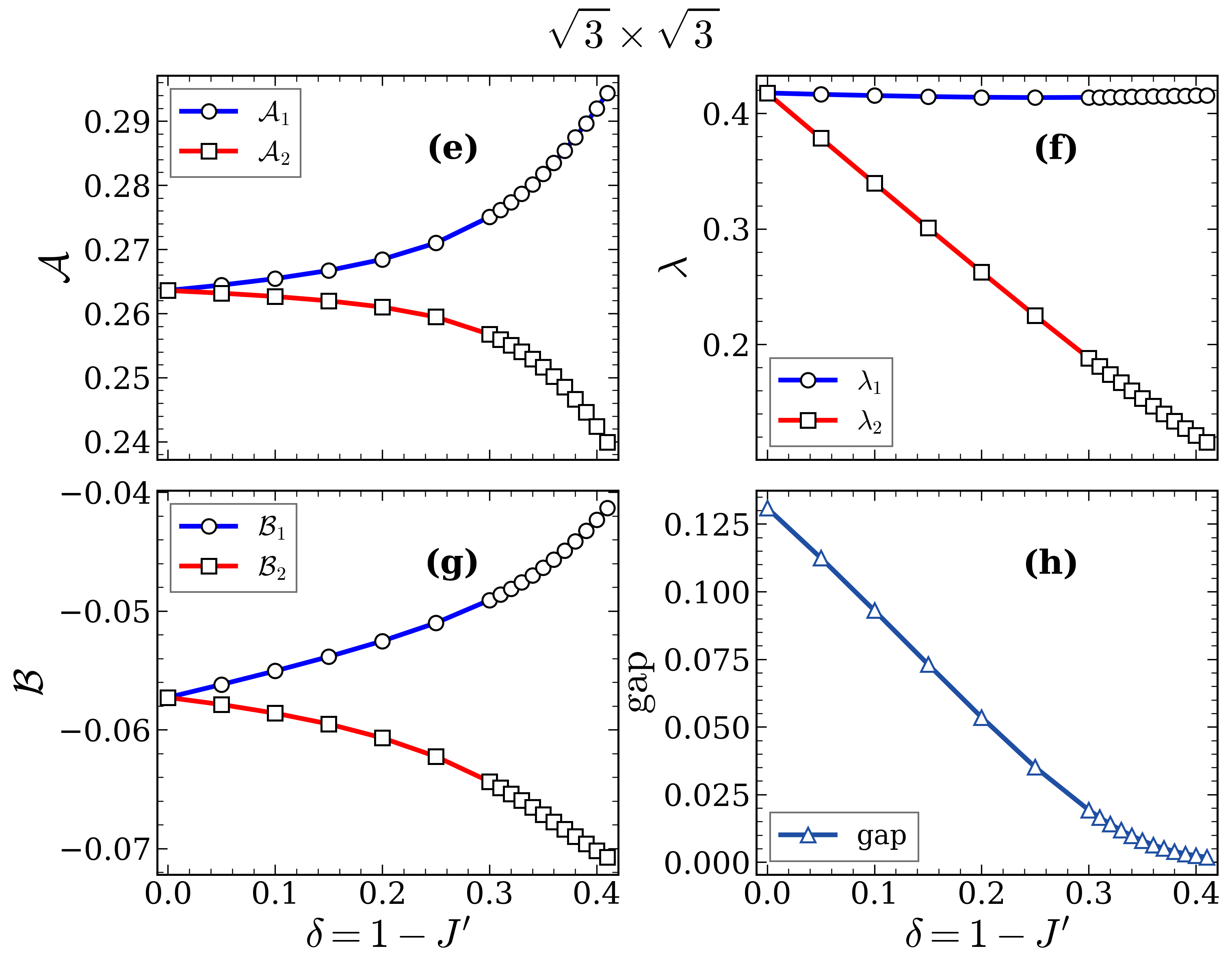}
\caption{
Evolution of the self-consistent SBMFT parameters with anisotropy
$\delta=1-J'/J$.
(a)--(d) Results for the $q=0$ ansatz: pairing fields
$\mathcal A_{1,2}$, Lagrange multipliers $\lambda_{1,3}$,
hopping fields $\mathcal B_{1,2}$, and the minimum spinon gap.
(e)--(h) Corresponding results for the $\sqrt{3}\times\sqrt{3}$ ansatz.
}
\label{fig:meanfields_gap}
\end{figure*}
%========================================

\subsection{Evolution of the mean-field parameters}

Figure~\ref{fig:meanfields_gap} summarizes the evolution of the self-consistent mean-field parameters as a function of anisotropy for the two ansätzes considered in this work. At $\delta=0$, both ansätzes correctly recover the isotropic kagome limit, with $\mathcal A_1 = \mathcal A_2$, $\mathcal B_1 = \mathcal B_2$, and $\lambda_1 = \lambda_2$ within numerical accuracy. This serves as an important consistency check, confirming that the anisotropic formulation smoothly connects to the well-established isotropic solution.

Upon increasing $\delta$, the anisotropy immediately lifts the equivalence between the two bond classes and the associated sublattices. For the $q=0$ ansatz, this splitting remains relatively weak over the range of parameters studied. The pairing amplitudes $\mathcal A_1$ and $\mathcal A_2$ evolve smoothly and in opposite directions, with $\mathcal A_1$ showing a slight enhancement while $\mathcal A_2$ is mildly suppressed. The hopping parameters exhibit a similarly modest reconstruction, with $\mathcal B_1$ becoming slightly more negative and $\mathcal B_2$ slightly less negative. In contrast, the Lagrange multipliers show a much more pronounced response: while $\lambda_1$ remains nearly constant, $\lambda_2$ decreases significantly with increasing $\delta$. This behavior reflects the increasing imbalance in local boson density constraints induced by the anisotropic exchange network, and signals a strong redistribution of low-energy weight across inequivalent sublattices.

%===========================================
\begin{figure*}[t!]
\includegraphics[width=17.2cm,height=6cm]{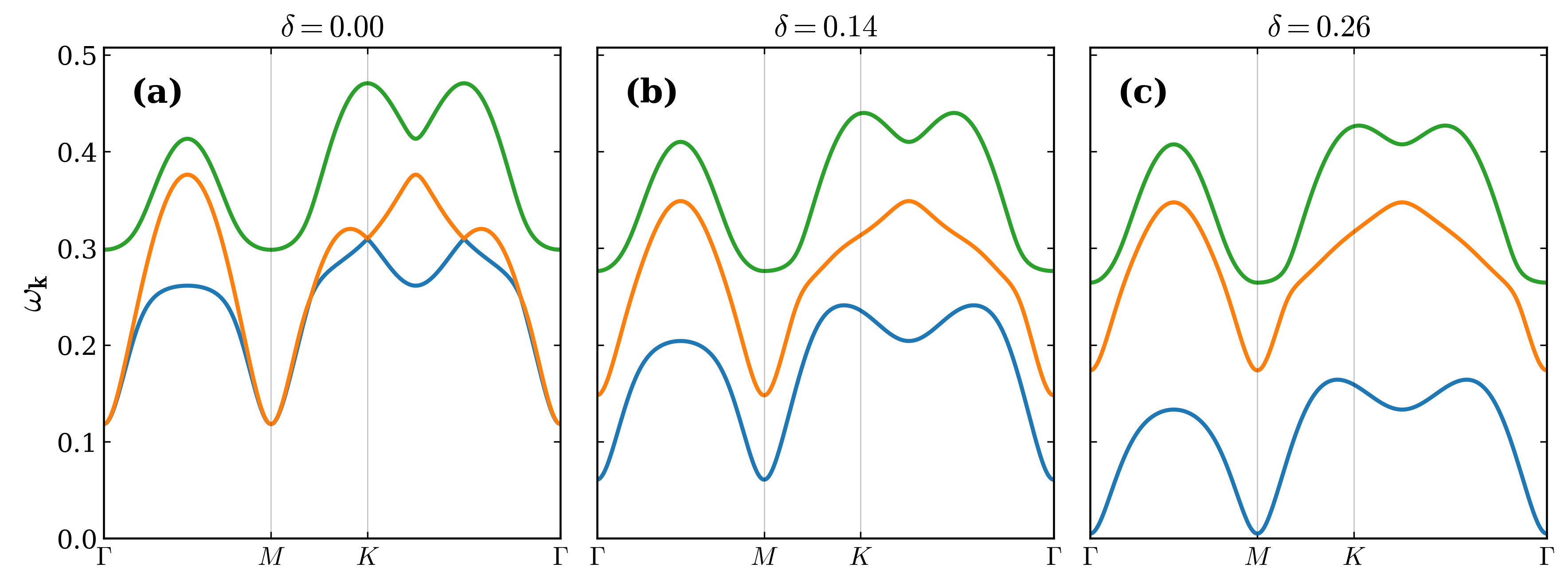}
\includegraphics[width=17.2cm,height=6cm]{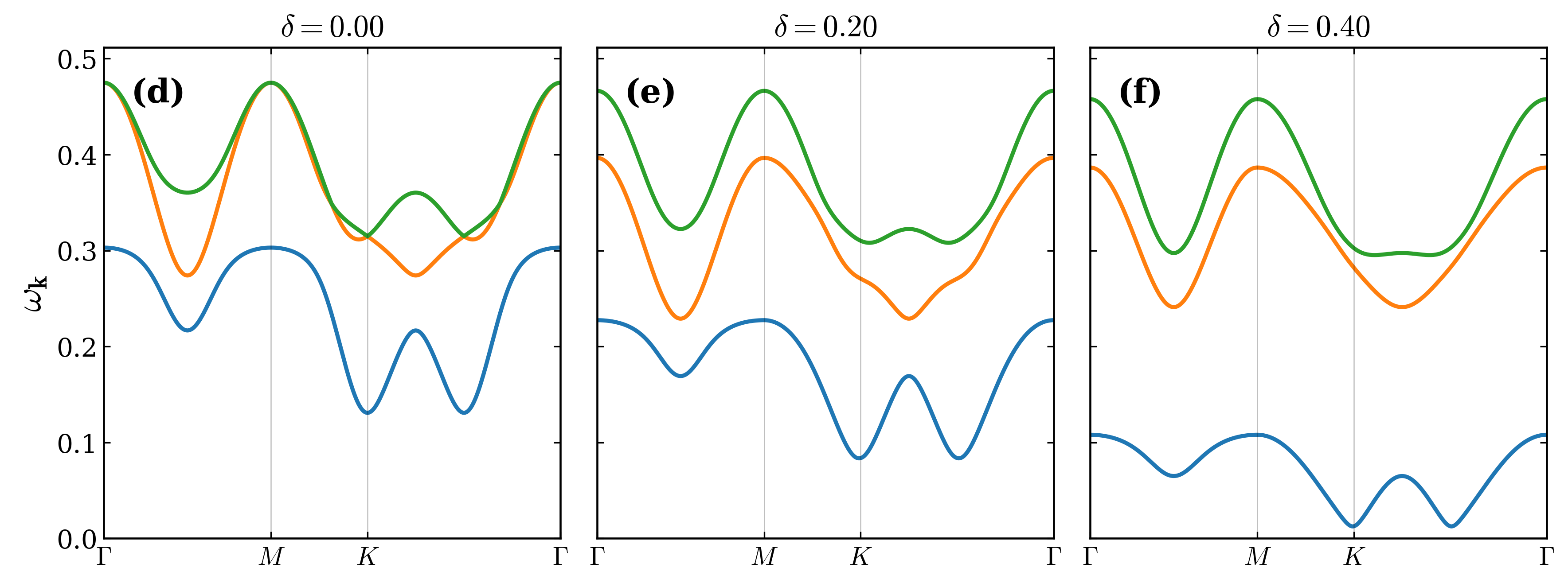}
\caption{
Spinon dispersions along the high-symmetry path
$\Gamma$--$M$--$K$--$\Gamma$ for representative anisotropies.
(a-c) The $q=0$ ansatz for $\delta=0.00,0.14,0.26$.
(d-f) The $\sqrt{3}\times\sqrt{3}$ ansatz for
$\delta=0.00,0.20,0.40$.
The softening of the lowest band tracks the reduction of the spinon gap with increasing anisotropy.
}
\label{fig:spinon_dispersion}
\end{figure*}
%========================================

The $\sqrt{3}\times\sqrt{3}$ ansatz displays a qualitatively similar but quantitatively much stronger sensitivity to anisotropy. In this case, the separation between $\mathcal A_1$ and $\mathcal A_2$ becomes substantial, with $\mathcal A_1$ increasing steadily and $\mathcal A_2$ decreasing more rapidly as $\delta$ grows. The hopping amplitudes also undergo a more pronounced reorganization compared to the $q=0$ state, with $\mathcal B_1$ becoming less negative while $\mathcal B_2$ decreases further. As in the $q=0$ case, the Lagrange multipliers split strongly, with $\lambda_2$ exhibiting a rapid downward renormalization. These trends indicate that the $\sqrt{3}\times\sqrt{3}$ saddle point is more strongly reshaped by anisotropy, reflecting its greater sensitivity to the underlying bond geometry and PSG structure.

A key quantity that captures the physical consequences of these parameter changes is the spinon gap. For both ansätzes, the minimum spinon gap decreases monotonically with increasing $\delta$ and eventually approaches zero. However, the rate at which the gap closes differs significantly between the two cases: the $q=0$ ansatz softens at relatively small anisotropy, whereas the $\sqrt{3}\times\sqrt{3}$ ansatz remains gapped over a broader parameter range. Within SBMFT, the closing of the spinon gap signals an instability toward spinon condensation, which corresponds to the onset of magnetic order at the mean-field level. In the present work, we focus exclusively on the gapped regime and do not attempt to continue the solution into the condensed phase.

%===========================================
\begin{figure*}[t!]
\includegraphics[width=18cm,height=11cm]{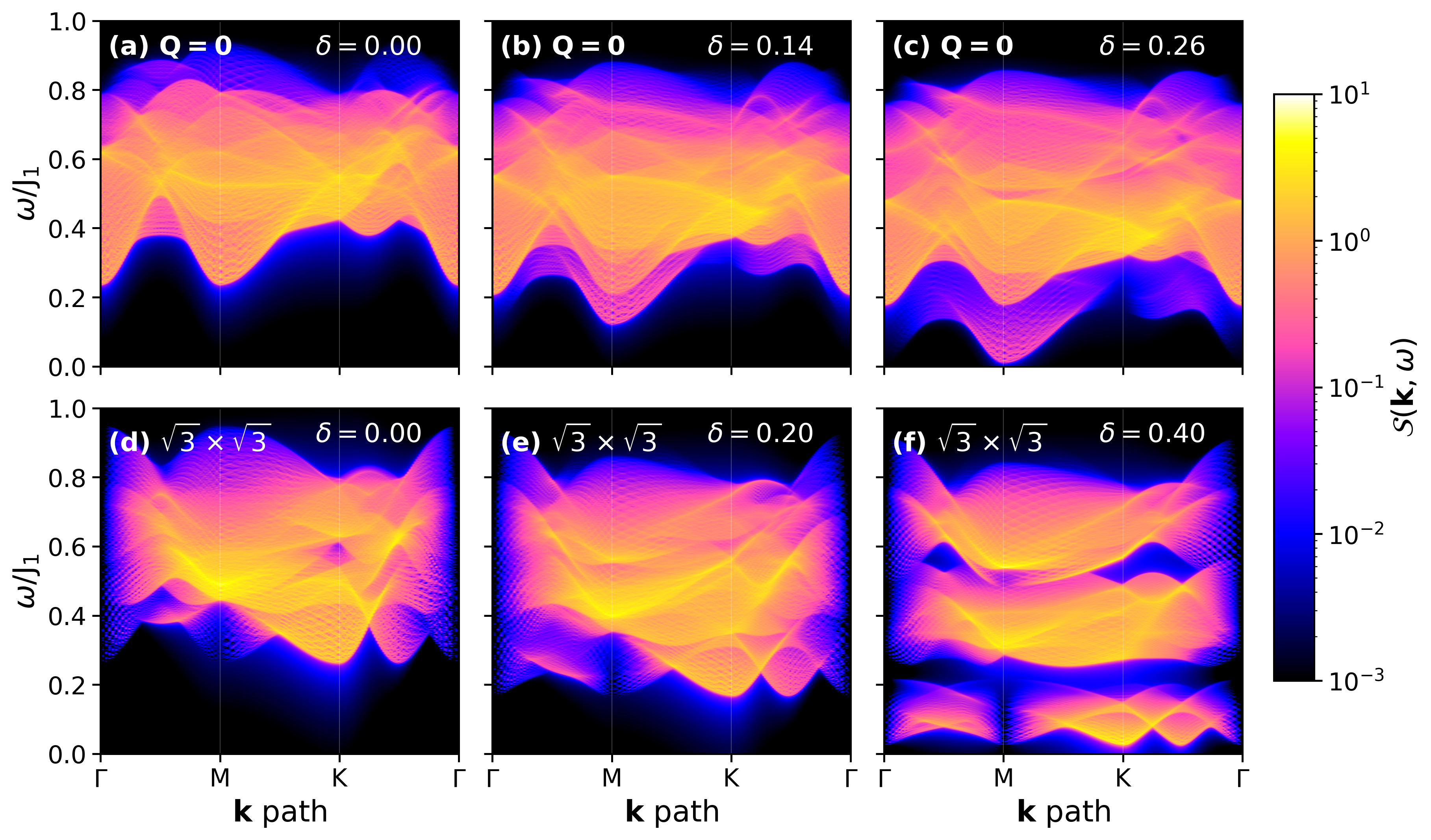}
\caption{
Zero-temperature dynamical structure factor $S({\bf k},\omega)$ along
$\Gamma$--$M$--$K$--$\Gamma$ for representative anisotropies.
(a)--(c) Results for the $q=0$ ansatz at
$\delta=0.00,0.14,0.26$.
(d)--(f) Results for the $\sqrt{3}\times\sqrt{3}$ ansatz at
$\delta=0.00,0.20,0.40$.
The intensity is shown on a logarithmic color scale.
}
\label{fig:dsf}
\end{figure*}
%========================================

\subsection{Spinon dispersions}

To gain further insight into the mechanism of gap closing, Fig.~\ref{fig:spinon_dispersion} shows the spinon dispersions along the high-symmetry path $\Gamma$--$M$--$K$--$\Gamma$ for representative values of $\delta$. The three bands arise from the three sublattices in the kagome unit cell.

At the isotropic point, both ansätzes exhibit fully gapped spectra with well-separated bands. As anisotropy is introduced, the dominant effect is a strong renormalization of the lowest spinon band, while the higher-energy bands remain comparatively rigid. Their bandwidths and overall energy scales are only weakly modified, indicating that the high-energy sector of the mean-field spectrum is relatively insensitive to moderate anisotropy.

In contrast, the lowest band undergoes a pronounced downward shift and develops momentum-dependent softening. For the $q=0$ ansatz, the softening occurs most prominently near the $\Gamma$ and $M$ regions of the Brillouin zone as $\delta$ increases toward $0.26$. The instability is therefore controlled by specific low-energy spinon modes, whose momenta determine the candidate ordering wavevectors after condensation.

For the $\sqrt{3}\times\sqrt{3}$ ansatz, a similar qualitative evolution occurs over a larger anisotropy window. As $\delta$ increases up to $0.40$, the lowest band is strongly suppressed while the upper bands remain at relatively high energies. In this case, the softening is associated with modes near the $K$ region and an intermediate momentum $K_1$ along the $\Gamma$--$K$ direction, with $K_1\simeq K/2$. Thus, the $\pi$-flux ansatz develops a distinct momentum-space pattern of low-energy modes compared with the $q=0$ ansatz.

This selective softening of the lowest spinon branch directly controls the low-energy onset of the dynamical response, since the lower edge of the two-spinon continuum is determined by sums of single-spinon energies. The dispersion results therefore identify both the closing of the spinon gap and the momenta of the soft modes that enter the incipient spinon-condensation instability.

\subsection{Dynamical structure factor}

The consequences of the spinon spectrum for experimentally observable
quantities are illustrated in Fig.~\ref{fig:dsf}, which shows the
zero-temperature dynamical structure factor $S({\bf k},\omega)$ along the
high-symmetry path $\Gamma$--$M$--$K$--$\Gamma$. At $T=0$, the Schwinger-boson
ground state contains no thermally excited spinons, and the response arises
from the creation of pairs of spinons. The lower edge of the two-spinon
continuum is therefore given by
\begin{equation}
    \omega_{\rm edge}({\bf k})
    =
    \min_{{\bf q},m,n}
    \left[
        \omega_m(-{\bf q})
        +
        \omega_n({\bf k}+{\bf q})
    \right] .
    \label{eq:continuum_edge}
\end{equation}
Thus, the low-energy onset of $S({\bf k},\omega)$ is controlled not only by
the minimum value of the spinon gap, but also by the momenta at which the
lowest spinon band softens. In particular, if low-energy spinons occur near
momenta ${\bf k}_0$ and symmetry-related partners, the two-spinon response is
enhanced at total momenta obtained by combining such soft modes, modulo a
reciprocal lattice vector.

For the $q=0$ ansatz, the isotropic system exhibits a broad and fully gapped
two-spinon continuum. Upon increasing anisotropy, the lower edge of the
continuum shifts downward, reflecting the softening of the lowest spinon band
seen in Fig.~\ref{fig:spinon_dispersion}. The most pronounced low-energy
spectral weight develops near the $\Gamma$ and $M$ regions of the path. This is
consistent with the corresponding single-spinon band structure, where the
softest modes occur near the same representative momenta although the low-energy spectral weight at the $\Gamma$ point is suppressed. Since the dynamical
structure factor probes two-spinon states, the observed low-energy features
should be interpreted as the continuum response associated with pairs of
soft spinons rather than as single-particle modes.

For the $\sqrt{3}\times\sqrt{3}$ ansatz, the anisotropy produces a similar
downward shift of the continuum edge, but the low-energy weight appears in a
different part of momentum space. In this case the softening is most visible
near the $K$ region and near the intermediate momentum $K_1$. This momentum dependence
reflects the different PSG structure and Bogoliubov coherence factors of the
$\pi$-flux ansatz. The resulting spectral-weight distribution is therefore not
a simple rescaling of the $q=0$ result, even though both ansätzes show the same
overall trend of a decreasing continuum threshold with increasing anisotropy.

Overall, the DSF confirms that anisotropy mainly reorganizes the low-energy spinon sector, lowering the continuum edge and concentrating spectral weight near the soft-mode momenta as the system approaches a spinon-condensation instability.

\section{Magnetic orders from soft-spinon modes}
\label{sec:soft_spinon_order}

The closing of the spinon gap within SBMFT signals an instability of the
gapped spin-liquid saddle point toward magnetic order. Although a fully
self-consistent treatment of the condensed phase requires supplementing the
mean-field equations with an explicit condensate density, the structure of the
incipient magnetic order can already be inferred from the eigenvector of the
soft spinon mode. We therefore use the lowest-energy Bogoliubov eigenmode as a
diagnostic of the magnetic texture selected by spinon condensation.

For each self-consistent mean-field solution, we first examine the lowest
positive spinon branch $\omega_1({\bf k})$ obtained from Eq.~\eqref{eq:bosonic_eig}
over the Brillouin zone. Let ${\bf k}_0$ denote a momentum at which this branch
becomes minimal. The corresponding positive-norm eigenvector is written as
\begin{equation}
    {\bf v}_{1{\bf k}_0}
    =
    \left(
    u_{u},u_{v},u_{w},
    v_{u},v_{v},v_{w}
    \right)^T ,
    \qquad
    {\bf v}_{1{\bf k}_0}^{\dagger}\tau_3{\bf v}_{1{\bf k}_0}=1 ,
    \label{eq:soft_mode_eigenvector}
\end{equation}
where the first three components belong to the particle sector and the last
three to the hole sector in the Nambu basis introduced above. The sublattice
index is denoted by $s=u,v,w$.

%===========================================
\begin{figure*}[t!]
    \centering
    \includegraphics[width=0.98\textwidth]{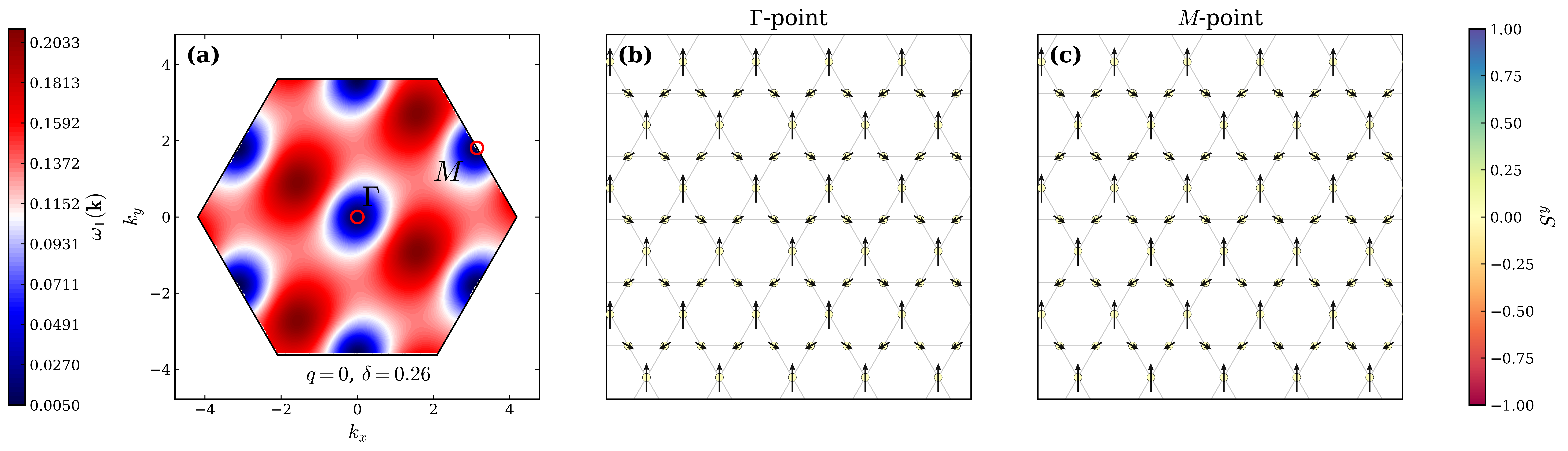}
    \includegraphics[width=0.98\textwidth]{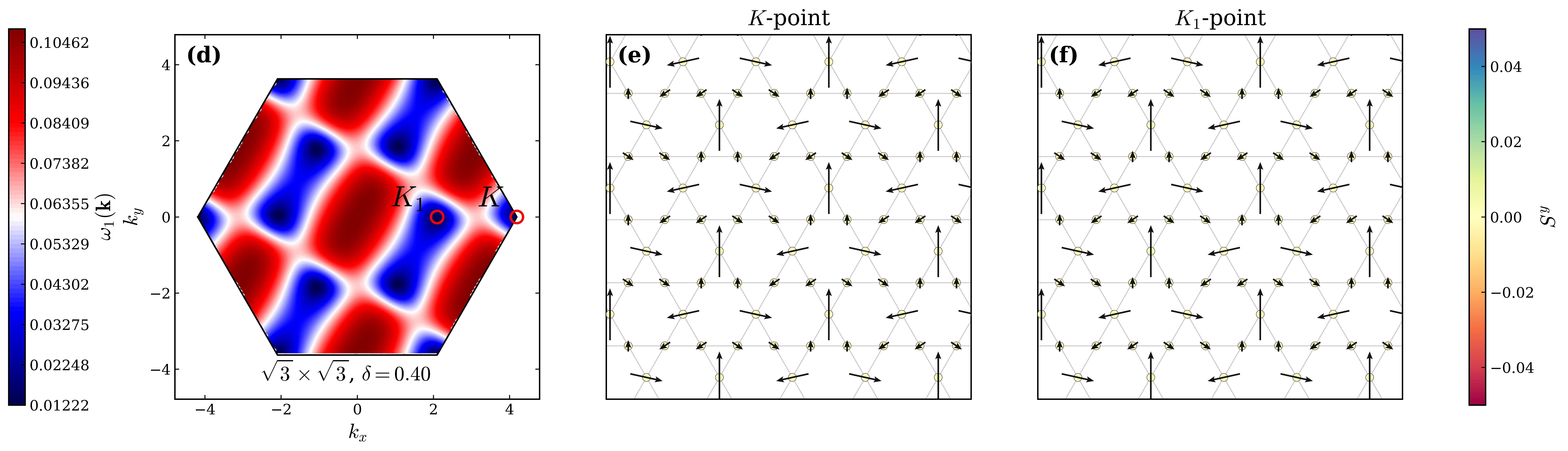}
    \caption{
    Soft-spinon modes and reconstructed spin textures for the anisotropic kagome antiferromagnet.
    (a--c) Results for the $q=0$ ansatz at $\delta=0.26$: lowest spinon band $\omega_1({\bf k})$ and condensate textures at $\Gamma$ and $M$.
    (d--f) Results for the $\sqrt{3}\times\sqrt{3}$ ansatz at $\delta=0.40$: lowest spinon band and condensate textures at $K$ and $K_1$, where $K_1$ lies along the $\Gamma$--$K$ direction.
    Red circles mark the selected soft momenta. Arrows show $(S^x,S^z)$ and the color scale shows $S^y$. Strong bonds are along $x$-direction.
    }
    \label{fig:soft_spinon}
\end{figure*}
%===========================================

To construct a real-space condensate texture, we associate the soft-mode
eigenvector with a local Schwinger-boson spinor on sublattice $s$,
\begin{equation}
    z_s({\bf R})
    =
    e^{i{\bf k}_0\cdot{\bf R}}
    \begin{pmatrix}
        u_s({\bf k}_0) \\
        v_s^{*}({\bf k}_0)
    \end{pmatrix}
    +
    e^{-i{\bf k}_0\cdot{\bf R}}
    \begin{pmatrix}
        u_s(-{\bf k}_0) \\
        v_s^{*}(-{\bf k}_0)
    \end{pmatrix}.
    \label{eq:condensate_spinor}
\end{equation}
Here ${\bf R}$ is a Bravais lattice vector, and the two terms represent
condensation of the symmetry-related modes at $\pm{\bf k}_0$. Including both
momenta produces a real-space spin configuration rather than a single complex
plane-wave condensate.

The corresponding spin expectation value is reconstructed using the Schwinger-boson representation,
\begin{equation}
    {\bf S}_s({\bf R}) = \frac{1}{2} z_s^\dagger({\bf R}) {\boldsymbol\sigma} z_s({\bf R}) .
    \label{eq:spin_from_condensate}
\end{equation}
This construction provides the spin texture associated with condensation of the softest spinon mode, and thus serves as a useful indicator of the magnetic order proximate to a given spin-liquid saddle point.

In practice, we obtain representative real-space textures by evaluating Eq.~\eqref{eq:spin_from_condensate} using the lowest Bogoliubov eigenvectors, as shown in Fig.~\ref{fig:soft_spinon}. For the $q=0$ ansatz, we consider soft modes near $\Gamma$ and $M$ at $\delta=0.26$, while for the $\sqrt{3}\times\sqrt{3}$ ansatz we use modes near $K$ and the intermediate point $K_1$ along the $\Gamma$--$K$ direction at $\delta=0.40$. In the real-space plots, arrows indicate the in-plane spin components, and their lengths reflect the spatial variation of the ordered moment derived from the soft eigenvectors.

Because the physical spin operator is bilinear in Schwinger bosons, a condensate at momentum ${\bf k}_0$ produces magnetic order at wave vector ${\bf Q}=2{\bf k}_0$. This explains the close similarity between the $\Gamma$ and $M$ textures in the $q=0$ case, since $2M$ is equivalent to a reciprocal lattice vector. Likewise, in the $\sqrt{3}\times\sqrt{3}$ case, the $K_1$ mode corresponds to ordering at ${\bf Q}=K$, while the $K$ mode maps to another Brillouin-zone corner. The resulting patterns should therefore be interpreted as representative ordering tendencies associated with the soft modes, rather than fully optimized classical configurations.

For the $q=0$ ansatz, the reconstructed magnetic order is relatively simple but anisotropic. The ordered moments exhibit a clear separation between sites on the strong $J$-bond backbone and the remaining sites. In particular, spins on the strong bonds develop reduced moments, while the remaining sites—referred to as \emph{dangling spins}, which connect neighboring chains through the weaker $J'$ couplings—carry enhanced amplitudes. This redistribution reflects the hierarchy of exchange interactions: strong bonds favor local antiferromagnetic correlations and partially suppress moment formation, whereas the dangling spins, being less constrained, develop larger condensate weight. The resulting $q=0$ state is thus characterized by a spatially modulated moment profile that preserves the lattice periodicity.

This structure can be understood in light of the classical ground-state constraints. The Hamiltonian may be rewritten (up to a constant) as
\begin{equation}
\mathcal{H} = \frac{J}{2} \sum_{\triangle} \left( \mathbf{S}_1 + \mathbf{S}_2 + \alpha\, \mathbf{S}_d \right)^2,
\qquad \alpha = \frac{J'}{J},
\end{equation}
where $\mathbf{S}_1$ and $\mathbf{S}_2$ denote the spins on a strong bond and $\mathbf{S}_d$ the dangling spin on the same triangle. Minimization imposes the local constraint $\mathbf{S}_1 + \mathbf{S}_2 + \alpha\, \mathbf{S}_d = 0$. In the isotropic limit $\alpha=1$, this reduces to the familiar $120^\circ$ configuration. For $\alpha<1$, the unequal weighting modifies the balance between spins: the two spins on the strong bond tend to align more antiparallel, while the dangling spin adjusts to compensate. The $q=0$ order may thus be viewed as a quantum realization of a distorted $120^\circ$ configuration. Importantly, beyond this classical picture, quantum fluctuations further renormalize the moments, leading to a pronounced suppression along the strong bonds—consistent with incipient singlet formation—while leaving the dangling spins comparatively less affected.

The $\sqrt{3}\times\sqrt{3}$ ansatz leads to a more strongly modulated texture with an enlarged unit cell, as shown in Fig.~\ref{fig:soft_spinon}(e) and (f). As in the $q=0$ case, the moment distribution reflects the underlying exchange anisotropy, with reduced amplitudes on the strong bonds and enhanced moments on the dangling spins. However, the spin orientations now vary more significantly within the unit cell, producing several inequivalent local configurations and a more intricate noncollinear pattern.

In contrast to the $q=0$ case, the connection to the classical constraint is less direct. While the isotropic $\sqrt{3}\times\sqrt{3}$ state satisfies the $120^\circ$ condition uniformly, it is not clear that a simple deformation of this pattern can satisfy the modified constraint for $\alpha<1$ on all triangles simultaneously. This suggests that the classical $\sqrt{3}\times\sqrt{3}$ order may not smoothly extend to the anisotropic regime. Instead, the state obtained from the Schwinger-boson analysis may reflect a more intricate compromise, in which both angular rearrangement and spatial modulation of the ordered moments are required to partially accommodate the local constraints. As before, quantum fluctuations further enhance this effect by preferentially suppressing the moments on the strong bonds.

We emphasize that these configurations are not fully self-consistent ordered solutions. They are obtained by condensing the soft spinon eigenmodes of the non-condensed mean-field saddle point, and therefore only indicate the magnetic order selected at the onset of the instability. A complete description of the ordered phase would require solving the SBMFT equations with a finite condensate density and incorporating its feedback on the bond fields and local constraints.

\section{Discussion and Outlook}

We have employed Schwinger-boson mean-field theory to investigate how spatial exchange anisotropy reshapes candidate spin-liquid states of the kagome antiferromagnet. The reduced lattice symmetry relative to the isotropic problem leads to a refined PSG classification, from which we constructed the corresponding nearest-neighbor ans\"atze. We then focused on representative $0$- and $\pi$-flux states continuously connected to the conventional $q=0$ and $\sqrt{3}\times\sqrt{3}$ kagome ans\"atze. Our central result is that exchange anisotropy primarily reorganizes the low-energy spinon sector: as $J'/J$ is reduced from unity, the inequivalent bond amplitudes and sublattice constraint fields split, with a more pronounced reconstruction in the $\sqrt{3}\times\sqrt{3}$ ansatz. Despite these differences, both states exhibit a common trend in which the lowest spinon branch softens strongly while higher-energy bands remain comparatively less affected, leading to a collapse of the spinon gap and signaling an instability of the non-condensed spin-liquid saddle point.

This evolution is directly reflected in the dynamical response. With increasing anisotropy, the lower edge of the two-spinon continuum shifts to lower energies, and the low-energy spectral weight becomes concentrated near momenta associated with the soft spinon modes. For the $q=0$ ansatz, these occur near $\Gamma$ and $M$, while for the $\sqrt{3}\times\sqrt{3}$ ansatz they appear near $K$ and an intermediate point $K_1\simeq K/2$. The dynamical structure factor thus provides a complementary probe of the same instability encoded in the single-spinon spectrum.

To characterize the magnetic order emerging at the onset of this instability, we reconstructed real-space spin textures from the soft Bogoliubov eigenvectors. The resulting configurations exhibit pronounced anisotropy in both structure and moment distribution: spins on the strong bonds are suppressed, while the more weakly coupled \emph{dangling} spins carry enhanced amplitudes, reflecting the hierarchy of exchange interactions. For the $q=0$ ansatz, the order corresponds to a relatively simple anisotropic deformation of the classical $120^\circ$ configuration. In contrast, the $\sqrt{3}\times\sqrt{3}$ ansatz develops a more strongly modulated texture with an enlarged unit cell, suggesting that the corresponding classical pattern does not smoothly adapt to the anisotropic constraints. These textures should be viewed as indicative of the ordering tendencies at the instability; a fully self-consistent treatment with finite condensate density is required to determine the ordered phases quantitatively.

Several important directions remain for future work. Extending the present analysis to fully self-consistent condensed solutions would clarify the nature of the ordered phases and their phase boundaries. It would also be valuable to systematically compare the full set of anisotropic PSG ans\"atze and to assess the role of gauge fluctuations beyond mean field, which may qualitatively modify the low-energy physics near the instability. Complementary numerical approaches, such as tensor-network methods or large-scale DMRG, could provide nonperturbative benchmarks in the anisotropic regime. 

On the experimental side, our results suggest that exchange anisotropy should leave clear fingerprints in neutron-scattering spectra, particularly in the redistribution of low-energy spectral weight and the selection of ordering wave vectors. These considerations may be relevant to recent Ti$^{3+}$-based kagome antiferromagnets~\cite{Jiang2020}, where lattice distortions generate nonuniform exchange networks that can deviate substantially from the ideal kagome limit. For example, structural studies indicate that the magnetic lattice can be viewed as a set of coupled, anisotropic chains or ``chainsaw''-like motifs, rather than a simple two-parameter anisotropy~\cite{Goto2016,Jeschke2019,Shirakami2019,Thennakoon2025b}. While the exchange pattern in such materials is more complex than the minimal anisotropic model considered here, our results highlight general mechanisms---namely, the reconstruction of low-energy spinon modes and their condensation into anisotropic magnetic textures---that are expected to remain operative. An important direction for future work is therefore to extend Schwinger-boson approaches to these more realistic exchange networks, which could provide a microscopic framework for interpreting the directional excitation continua and unconventional magnetic responses observed in titanium-based kagome compounds.

\begin{acknowledgments}
This work was supported by the U.S. Department of Energy, Office of Science, Basic Energy Sciences, through DE-SC0026087.
\end{acknowledgments}

\appendix

\section{Derivation of the PSG constraints}

\label{app:psg}

For the projective-symmetry-group (PSG) analysis of the anisotropic kagome lattice, we follow the general framework of Ref.~\cite{wang2006}, adapting it to the reduced point-group symmetry of the present model. We use a triangular Bravais lattice with primitive vectors $\mb{a}_1$ and $\mb{a}_2$, separated by $120^\circ$, together with a three-site basis $\bs{\delta}_u = \mb{a}_2/2$, $\bs{\delta}_v = \mb{a}_1/2$, and $\bs{\delta}_w = (\mb{a}_1 + \mb{a}_2)/2$. Lattice sites are labeled as $(r_1,r_2)_p$ with $r_1,r_2\in\mathbb{Z}$ and sublattice index $p=u,v,w$. We consider an anisotropic nearest-neighbor model with coupling $J$ on the $u$-$w$ and $v$-$w$ bonds, and coupling $J'$ on the $u$-$v$ bonds.

The space group of the anisotropic kagome lattice is generated by translations $T_1,T_2$, a reflection $\sigma$, and a $180^\circ$ rotation $R_\pi$, which act as
\begin{align}
    T_1&:\ (r_1,r_2)_p\to(r_1+1,r_2)_p, \\
    T_2&:\ (r_1,r_2)_p\to(r_1,r_2+1)_p, \\
    \begin{split}
        \sigma&:\ (r_1,r_2)_u\to(r_2,r_1)_v, \\
        &\quad\,(r_1,r_2)_v\to(r_2,r_1)_u, \\
        &\quad\,(r_1,r_2)_w\to(r_2,r_1)_w,
    \end{split} \\
    \begin{split}
        R_\pi&:\ (r_1,r_2)_u\to(-r_1,-r_2-1)_u, \\
        &\quad\,(r_1,r_2)_v\to(-r_1-1,-r_2)_v, \\
        &\quad\,(r_1,r_2)_w\to(-r_1-1,-r_2-1)_w.
    \end{split}
\end{align}

Within Schwinger-boson mean-field theory, a symmetry operation need only leave the ansatz invariant up to a gauge transformation. We therefore associate to each generator $X\in\{T_1,T_2,\sigma,R_\pi\}$ a site-dependent gauge transformation
\begin{equation}
    G_X:\ b_{(r_1,r_2)_p\alpha}\to e^{i\phi_{X,p}(r_1,r_2)}b_{(r_1,r_2)_p\alpha}.
\end{equation}
The PSG generators are thus $G_X X$. Since the invariant gauge group (IGG) is $Z_2$, all algebraic relations of the space group are required to hold only up to a sign. Physically, these signs encode distinct patterns of emergent gauge flux compatible with the symmetry.

The relevant space-group relations are
\begin{align}
    T_1T_2 &= T_2T_1,\\
    \sigma^2 &= 1,\\
    R_\pi^2 &= 1,\\
    \sigma R_\pi &= R_\pi\sigma,\\
    \sigma T_1\sigma^{-1} &= T_2,\\
    \sigma T_2\sigma^{-1} &= T_1,\\
    R_\pi T_1R_\pi^{-1} &= T_1^{-1},\\
    R_\pi T_2R_\pi^{-1} &= T_2^{-1}.
\end{align}
These relations form a complete set, since any group element can be reduced to $\sigma^a R_\pi^b T_1^m T_2^n$. In the following, all phase equations are understood modulo $2\pi$, and we repeatedly use the conjugation rule
\begin{equation}
    \phi_{Y^{-1}G_XY,p}(r_1,r_2) = \phi_{X,Yp}\!\left(Y(r_1,r_2)\right).
\end{equation}

\subsection{Translation PSG}

We first fix the gauge associated with translations. Under a general gauge transformation
\begin{equation}
    b_{(r_1,r_2)_p\alpha}\to e^{i\theta_p(r_1,r_2)}b_{(r_1,r_2)_p\alpha},
\end{equation}
the PSG phases transform as
\begin{multline}
    \phi_{X,p}(r_1,r_2)\to \phi'_{X,p}(r_1,r_2) \\
    = \theta_p(r_1,r_2) + \phi_{X,p}(r_1,r_2) \\
    - \theta_{X^{-1}p}\!\left(X^{-1}(r_1,r_2)\right).
\end{multline}
This gauge freedom allows us to set
\begin{equation}
    \phi_{T_1,p}(r_1,r_2)=0 ,
\end{equation}
by choosing $\theta_p$ recursively along the $r_1$ direction. The remaining freedom, independent of $r_1$, is used to impose
\begin{equation}
    \phi_{T_2,p}(0,r_2)=0 .
\end{equation}

Imposing the commutation relation $T_1T_2=T_2T_1$ up to IGG gives
\begin{equation}
    \phi_{T_2,p}(r_1,r_2)=p_1\pi r_1,
\end{equation}
with $p_1=0,1$. This parameter distinguishes between trivial and $\pi$-flux states threading the elementary parallelogram of the Bravais lattice.

\subsection{Reflection PSG}

We next impose relations involving $\sigma$. Solving the translation-reflection constraints gives
\begin{equation}
    \phi_{\sigma,p}(r_1,r_2) = \phi_{\sigma,p}(0,0) + q_1\pi r_1 + q_2\pi r_2 + p_1\pi r_1r_2.
\end{equation}
The condition $\sigma^2=1$ imposes $q_1=q_2$ and constrains the constant terms. Using the remaining gauge freedom, we can eliminate the linear terms and choose a symmetric form. This yields
\begin{equation}
    \phi_{\sigma,p}(r_1,r_2) = \frac{p_2\pi}{2} + p_1\pi r_1r_2,
\end{equation}
with $p_2=0,1$. The parameter $p_2$ encodes the projective action of reflection on the spinons and distinguishes distinct symmetry fractionalization classes.

\subsection{Rotation PSG}

Finally, we consider the $180^\circ$ rotation $R_\pi$. The translation-rotation relations fix the spatial dependence of $\phi_{R_\pi,p}$, leading to
\begin{align}
    \phi_{R_\pi,p}(r_1,r_2)
    &= c_p + q_3\pi r_1 + q_4\pi r_2 \notag \\
    &\qquad + (\text{sublattice-dependent terms}) .
\end{align}
with different coefficients for $u,v,w$ sublattices.

Imposing $R_\pi^2=1$ constrains the constants $c_p$, while the commutation relation $\sigma R_\pi = R_\pi \sigma$ further requires
\begin{equation}
    q_3 = q_4 \equiv p_4,
\end{equation}
and relates the constants across sublattices. The resulting phases can be written as
\begin{align}
    \phi_{R_\pi,u}(r_1,r_2) &= \frac{(p_3+p_4)\pi}{2} + p_4\pi(r_1+r_2), \\
    \phi_{R_\pi,v}(r_1,r_2) &= \frac{(p_3+p_4)\pi}{2} + p_1\pi + p_4\pi r_1 + (p_4+p_1)\pi r_2, \\
    \phi_{R_\pi,w}(r_1,r_2) &= \frac{(p_3+p_1)\pi}{2} + p_4\pi r_1 + (p_4+p_1)\pi r_2,
\end{align}
where $p_3=0,1$ arises from the $R_\pi^2$ constraint.

\subsection{Summary of PSG solutions}

Collecting the results, the algebraic PSG is characterized by four independent $\mathbb{Z}_2$ parameters $p_1,p_2,p_3,p_4$. In the gauge fixed above,
%\begin{widetext}
%\begin{align}
%    \phi_{T_1,p}(r_1,r_2) &= 0, \\
%    \phi_{T_2,p}(r_1,r_2) &= p_1\pi r_1, \\
%    \phi_{\sigma,p}(r_1,r_2) &= \frac{p_2\pi}{2} + p_1\pi r_1r_2, \\    \phi_{R_\pi,u}(r_1,r_2) &= \frac{(p_3+p_4)\pi}{2} + p_4\pi(r_1+r_2), \\\phi_{R_\pi,v}(r_1,r_2) &= \frac{(p_3+p_4)\pi}{2} + p_1\pi + p_4\pi r_1 + (p_4+p_1)\pi r_2, \\\phi_{R_\pi,w}(r_1,r_2) &= \frac{(p_3+p_1)\pi}{2} + p_4\pi r_1 + (p_4+p_1)\pi r_2.
%\end{align}
%\end{widetext}
\begin{align}
    \phi_{T_1,p}(r_1,r_2)
    &= 0, \\
    \phi_{T_2,p}(r_1,r_2)
    &= p_1\pi r_1, \\
    \phi_{\sigma,p}(r_1,r_2)
    &= \frac{p_2\pi}{2} + p_1\pi r_1r_2, \\
    \phi_{R_\pi,u}(r_1,r_2)
    &= \frac{(p_3+p_4)\pi}{2}
    + p_4\pi(r_1+r_2), \\
    \phi_{R_\pi,v}(r_1,r_2)
    &= \frac{(p_3+p_4)\pi}{2} + p_1\pi
    + p_4\pi r_1 \notag \\
    &\qquad
    + (p_4+p_1)\pi r_2, \\
    \phi_{R_\pi,w}(r_1,r_2)
    &= \frac{(p_3+p_1)\pi}{2}
    + p_4\pi r_1 \notag \\
    &\qquad
    + (p_4+p_1)\pi r_2 .
\end{align}
Thus, at the algebraic level, the anisotropic kagome lattice admits $2^4=16$ distinct $Z_2$ PSGs. Compared to the isotropic kagome case, the enlarged set of solutions reflects the reduced point-group symmetry: replacing the sixfold rotation symmetry by a $180^\circ$ rotation relaxes the algebraic constraints and allows for an additional independent PSG parameter. In practice, further constraints from a specific mean-field ansatz (e.g., requiring nonvanishing amplitudes on particular bonds) will reduce this set and select the physically relevant states.

\bibliography{ref.bib}% Produces the bibliography via BibTeX.

\end{document}